\documentclass[
preprint,
superscriptaddress,
amsmath,
amssymb,
aps,
citeautoscript
]
{revtex4-2}
\usepackage{graphics, graphicx}
\usepackage[italic]{mathastext}
\usepackage{bm}
\usepackage{mathtools,amsmath}
\usepackage{adjustbox}
\usepackage{array,booktabs,tabularx}
\usepackage{siunitx}
\usepackage{upquote}
\usepackage{hhline}
\usepackage{stackengine}
\usepackage{xcolor}
\usepackage{setspace} 
\usepackage{hyperref}
\usepackage{microtype}
\usepackage[capitalize]{cleveref}
\usepackage{xr}
\usepackage{float}
\usepackage{xspace}
\usepackage{xcolor}
\usepackage{longtable}
\usepackage{supertabular}
\usepackage{braket}

\hypersetup{
    colorlinks,
    linkcolor={blue!50!black},
    citecolor={blue!50!black},
    urlcolor={blue!80!black}
}

\newcolumntype{C}[1]{>{\centering\arraybackslash}p{#1}}

\newcommand{\xyz}{$\sigma_{xy}^z$\xspace} 
\newcommand{\xyx}{$\sigma_{xy}^x$\xspace} 
\newcommand{\xyy}{$\sigma_{xy}^y$\xspace} 
\newcommand{\xx}{$\sigma_{xx}$\xspace} 
\newcommand{\units}{$e/4\pi$\xspace} 
\newcommand{\unitqs}{$e/2\pi$\xspace} 
\newcommand{\ef}{Fermi energy\xspace} 
\newcommand{\YCI}{Y$_2$C$_2$I$_2$\xspace}
\newcommand{\TS}{Ta$_4$Se$_2$\xspace}
\newcommand{\YB}{Y$_2$Br$_2$\xspace}
\newcommand{\CLA}{CuLi$_2$As\xspace} 
\newcommand{\SnP}{Sn$_2$P\xspace} 
\newcommand{\frs}{F.R.S.\xspace}

\newcommand{\figpath}{Figs}

\begin{document}

\begin{abstract}
  Spin Hall effect (SHE) in two-dimensional (2D) materials 
  is promising  
  to effectively manipulate spin angular momentum 
  and identify topological properties.
  In this work, 
  we implemented an automated Wannierization with spin-orbit coupling on 426 non-magnetic monolayers
  including 210 metal and 216 insulators. 
  Intrinsic spin Hall conductivity (SHC) 
  has been calculated   
  to find candidates exhibiting novel properties.
  We discover that \YCI has an unconventional SHE with canted spin 
  due to low crystal symmetry,
  \TS is a metallic monolayer with exceptionally high SHC,  
  and the semi-metal \YB possesses efficient charge-to-spin conversion
  induced by anti-crossing in bands. 
  Moreover,
  quantum spin Hall insulators are investigated for  quantized SHC. 
  The present work 
  provides a high-quality   Wannier Hamiltonian database of  2D materials,
  and 
  paves the way for the integration of 2D materials into
  high-performance and low-power-consumption spintronic devices. 
\end{abstract}

\title{High-throughput calculations of spin Hall conductivity in non-magnetic 2D materials}

\author{Jiaqi Zhou} \email{jiaqi.zhou@uclouvain.be}
\affiliation{Institute of Condensed Matter and Nanosciences, Université catholique de Louvain, 1348 Louvain-la-Neuve, Belgium}

\author{Samuel Poncé}
\email{samuel.ponce@uclouvain.be}
\affiliation{European Theoretical Spectroscopy Facility and Institute of Condensed Matter and Nanosciences, Université catholique de Louvain, 1348 Louvain-la-Neuve, Belgium}
\affiliation{WEL Research Institute, Avenue Pasteur 6, 1300 Wavre, Belgium}

\author{Jean-Christophe Charlier} \email{jean-christophe.charlier@uclouvain.be}
\affiliation{Institute of Condensed Matter and Nanosciences, Université catholique de Louvain, 1348 Louvain-la-Neuve, Belgium}

\date{\today} \maketitle

\clearpage

Spin Hall effect (SHE) is a phenomenon 
emerging from spin-orbit coupling (SOC) 
where an external electric field can induce a transverse spin current~\cite{Sinova2015Oct}.
As an effective method 
to control spin transport, 
the SHE has been extensively  investigated in bulk heavy metals~\cite{Tao2018Jun}.
However, 
simple heavy metals are constrained by symmetry and  
only present conventional SHE where 
spin current, electric field, 
and spin orientation are orthogonal
to each other. 
Constrained by this strict rule,  
an external magnetic field is required to switch perpendicular magnetization~\cite{Liu2012Aug}, 
hindering device minimization. 
Besides, 
defined as the ratio of spin Hall conductivity (SHC)
to charge conductivity, 
the spin Hall ratio (SHR)
is reduced by   high charge conductivity in heavy metals, 
leading to a low  charge-to-spin conversion efficiency~\cite{Tao2018Jun}. 
Van der Waals materials emerge as promising candidates
to overcome these challenges. 
SHCs
have been studied in two-dimensional (2D)
transition-metal dichalcogenide (TMDC) 
materials~\cite{Ahn2020Jun,Safeer2019Feb,MacNeill2017Mar,Shi2019Oct,Guimaraes2018Feb,Stiehl2019Nov,Xu2020Apr,Mishra2024Nov}.
Compared with high-symmetry bulk heavy metals, 
2D materials with low-symmetry structures 
can induce unconventional SHEs~\cite{Zhou2021Nov,Ji2024Nov}.
In addition, the reduced charge conductivity in 
2D materials enables efficient charge-to-spin conversion~\cite{Vila2021Dec,Safeer2019Dec}. 
More importantly, 
the reduced dimension 
induces quantum spin Hall effect  
 in topological 2D materials~\cite{Grassano2023Sep}, 
offering robust and quantized SHCs in  bandgaps 
of quantum spin Hall insulators  (QSHIs)~\cite{Bai2022May, Matusalem2019Dec,Matthes2016Aug}. 
However, 
current researches on SHE of 2D materials are mostly limited to TMDC materials, 
while a  large-scale study on 2D materials for spintronics
is still absent.

Wannier functions~\cite{Marzari2012Oct,Marrazzo2023Dec}
are extensively used to study the SHC~\cite{Qiao2018Dec}. 
Recent works have developed frameworks
for the high-throughput Wannierization.
Symmetrized Wannier Hamiltonians
have been generated 
for the strained III-V semiconductor materials
by applying post-processing symmetrization~\cite{Gresch2018Oct}. 
In addition, the single-criterion decision-making algorithm
with no need for an initial guess 
has been validated for band interpolations
on a dataset of 200 materials~\cite{Vitale2020Jun}.
Moreover, 
Wannier Hamiltonians including SOC for 1406 3D and 365 2D materials
were generated using hydrogenic orbital projections
with hints from pseudopotentials~\cite{Garrity2021Apr}.
Recently, 
the projectability-disentangled Wannier functions
have been proposed 
where 
the initial guess is given by 
a projectability measure for each Bloch state 
onto atomic orbitals~\cite{Qiao2023Oct,Qiao2023Nov}.

In this work, we perform a high-throughput study of 
 spin Hall conductivity in non-magnetic 2D materials
 using density functional theory (DFT) and Wannier functions.
 Leveraging the scalar-relativistic \textsc{MC2D} database~\cite{MC2D1,MC2D2}, all the rare-earth-free
monolayers with up to 6 atoms per unit cell are considered, yielding
  426 exfoliable materials. 
These materials are investigated with 
$ab~initio$ calculations including SOC. 
Intrinsic SHCs are calculated in all the materials
to reveal exotic SHE. 
Unconventional SHE is found in a low-symmetry metal, 
materials with high SHC values are listed, 
and a 
high SHR is found in a semi-metal. 
Finally, SHCs of topological insulators
are studied to understand 
the importance of spin conservation for quantized SHC.

\section*{Results and discussions} 

To match the typical experimental setup,
we focus on the SHC where the spin current is along $x$, 
the electric field  is along $y$, 
and the spin orientation is along $\alpha$ direction. 
The  intrinsic SHC in a weak scattering limit can be 
  calculated by the Kubo
  formula~\cite{Qiao2018Dec,Zhang2021Oct}: 

\begin{equation} 
\label{eq:kubo_shc} 
  \sigma_{xy}^{\alpha} 
    = \frac{\hbar}{2e} \frac{e^2}{\hbar} \int_\text{BZ} \frac{\mathrm{d}^2
      \mathbf{k}}{(2\pi)^2} \Omega_{xy}^{\alpha}(\mathbf{k}), 
\end{equation} 
where BZ denotes the Brillouin zone, and the spin Berry curvature (SBC) is given as 
\begin{align}
  \Omega_{xy}^{\alpha}(\mathbf{k}) &= \sum_{n} f_{n\mathbf{k}}\Omega_{n, xy}^{\alpha}(\mathbf{k}), \label{eq:spinberry} \\
  \Omega_{n, xy}^{\alpha}(\mathbf{k}) &= \hbar^2 \sum_{m \ne n}
      \frac{-2\operatorname{Im}[\langle n\mathbf{k}| \hat{j}_{\alpha} |m\mathbf{k}\rangle
          \langle m\mathbf{k}|
          \hat{v}_{y}|n\mathbf{k}\rangle]}{(\varepsilon_{n\mathbf{k}}-\varepsilon_{m\mathbf{k}})^2
        + \eta^2}. 
\label{eq:bandspinberry}
\end{align}
\noindent
In the above equations, 
$f_{n{\mathbf{k}}}$ 
denotes the Fermi-Dirac distribution 
$ f_{n{\mathbf{k}}}(\mu, T) = \frac{1}{{e^{(\varepsilon_{n\mathbf{k}} - \mu)/(k_B T)} + 1}}$,
where
$\mu$ is the Fermi energy.
 $\hat{j}_{\alpha} = \frac{1}{2} \{\hat{\sigma}_{\alpha}
    \hat{v}_x + \hat{v}_x \hat{\sigma}_{\alpha} \}$ is the spin current operator,
  $\hat{\sigma}_{\alpha} $ is the Pauli operator, $ \hat{v}_x$ and $ \hat{v}_y$ are
  velocity operators, 
  $\eta$ is the broadening approximated as extrinsic scattering~\cite{Li2015Apr}. 

To compute the SHC on ultra-dense $\mathbf{k}$ grids, 
we construct an 
$ab~initio$ Wannier Hamiltonian database
and assess its quality by comparing the average  
differences between DFT bands and Wannier bands,
given as~\cite{Qiao2023Oct,Qiao2023Nov} \begin{equation}
    \label{eq:distance}
    \eta_{\nu}=\sqrt{\frac{\sum_{n \mathbf{k}} \tilde{f}_{n
    \mathbf{k}}\left(\varepsilon_{n \mathbf{k}}^{\mathrm{DFT}}-\varepsilon_{n
      \mathbf{k}}^{\mathrm{Wan}}\right)^2}{\sum_{n \mathbf{k}} \tilde{f}_{n
    \mathbf{k}}}} \end{equation} 
where $\bf k$ denotes eigenstates along the
  high-symmetry $\bf k$-path of the band structure, $\tilde{f}_{n
    \mathbf{k}}=\sqrt{f_{n
          \mathbf{k}}^{\mathrm{DFT}}\left(\mathrm{E}_{\mathrm{F}}+{\nu}, T\right) f_{n
          \mathbf{k}}^{\mathrm{Wan}}\left(\mathrm{E}_{\mathrm{F}}+{\nu}, T\right)}$ 
where $\mathrm{E}_{\mathrm{F}}+{\nu}$ represents energy position, 
thus $\eta_{\nu}$ denotes the
 difference given by comparing
 $\varepsilon_{n \mathbf{k}}^{\mathrm{DFT}}$
and $\varepsilon_{n \mathbf{k}}^{\mathrm{Wan}}$ below $\mathrm{E}_{\mathrm{F}}+{\nu}$.
Since the Wannier Hamiltonian describes the low-energy subspace, it is
  expected that the Wannier-interpolated bands will deviate more from the DFT results in the higher
  energy region.
Therefore, the higher ${\nu}$ is, the larger $\eta_{\nu}$ is expected to be.
The symbols $\eta_1$ and $\eta_2$ denote the band distance for the bands below
  $\mathrm{E}_{\mathrm{F}}+1~\mathrm{eV}$ and
  $\mathrm{E}_{\mathrm{F}}+2~\mathrm{eV}$, respectively.
For $\eta_1$, 98\% 
  of materials show a  band distance below 10~meV, and 92\% 
  of materials show a   small band distance below 2~meV.
For $\eta_2$, 93\% 
  of materials show a distance below 10~meV, and 70\% 
   of materials show a   distance below 2~meV,
  highlighting the overall quality of Wannier Hamiltonians in our
  database.
$\eta$ distributions 
  are provided in Supplementary Fig. 1(b).
Bands of all the metals and insulators
are present in Supplementary Secs. 4 and 5.

\subsection*{Unconventional spin Hall effect}

\begin{figure}[tb]
  \includegraphics{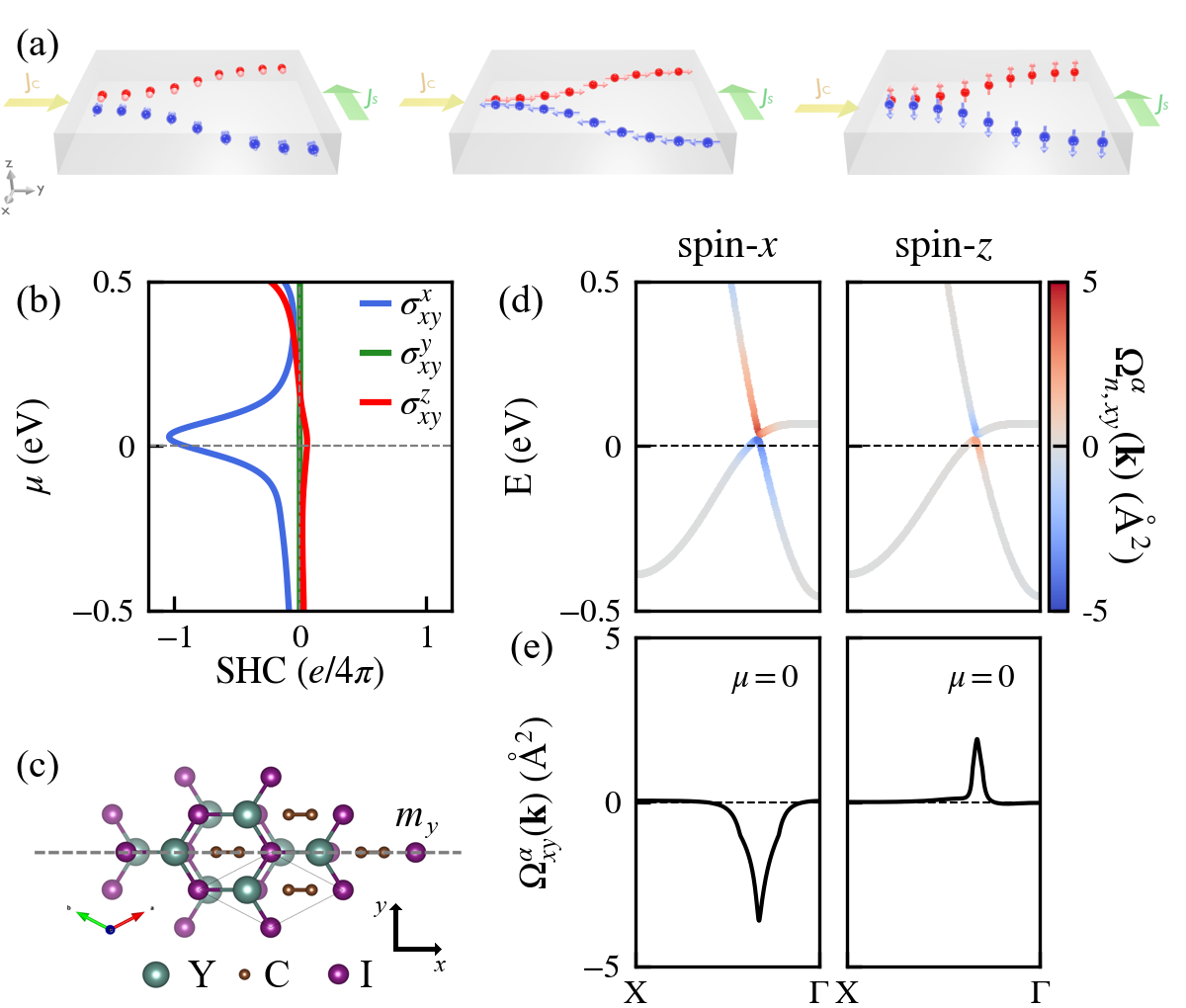}
  \caption{
  \textbf{Canted spin current in \YCI.}
  (a) Diagram of spin Hall effect components  \xyx, \xyy, and \xyz, respectively.
  (b) \ef $\mu$-dependent room-temperature spin Hall conductivity components,
  (c) atomic structure and mirror symmetry, 
  (d) band-resolved spin Berry curvature defined by \cref{eq:bandspinberry},
  and 
  (e) $\mathbf{k}$-resolved spin Berry curvature defined by \cref{eq:spinberry}
  for \YCI~[C2/m] with unconventional spin Hall effect. 
  }
  \label{fig:Fig1}  
\end{figure}

Neumann's principle states that  
symmetries of the physical property must 
include all  symmetries of the crystal~\cite{Seemann2015Oct}, 
indicating that  a high-symmetry crystal   
would impose more restrictions on the allowed property.  
As a result, 
  high-symmetry monolayers can only present one unique non-zero
SHE component \xyz~\cite{Roy2022Apr}, 
which is called conventional SHE
where the spin orientation is 
perpendicular to the $xy$-plane as shown in \cref{fig:Fig1}(a).
In contrast, low-symmetry materials can lead to 
unconventional SHE~\cite{Roy2022Apr}. 
The  unconventional SHE 
is important for the field-free switching 
of perpendicular magnetization, 
since it can provide a canted torque 
to trigger the spin dynamics 
without external magnetic field~\cite{Liu2021Mar}.

To discover intrinsically  unconventional SHE, 
we focus on metals with non-zero
\xyx and \xyy as illustrated in  \cref{fig:Fig1}(a).
A symmetry precision of 0.01 was applied 
in the {\sc{ase}} package~\cite{Larsen2017Jun}
to screen materials with low-symmetry space group.
After screening,
we find 17 metals with low symmetry. 
Most of them present small SHCs. 
Notably,
\YCI is found to present a high $\sigma_{xy}^x$ 
at the \ef as shown in \cref{fig:Fig1}(b).  
With the {C2/m} space group, 
\YCI only has one mirror symmetry $m_y$
as shown in \cref{fig:Fig1}(c), 
allowing the spin current with spin orientation in $xz$-plane. 
To analyze  the
$\sigma_{xy}^x$ and  $\sigma_{xy}^z$ behaviors,
\cref{fig:Fig1}(d) presents the bands projected by SBC
along the X-$\Gamma$ $\bf k$-path.
The SOC results in  an anti-crossing gap  in  \YCI. 
The positive and negative $\Omega_{n, xy}^x$ 
are perfectly separated by the \ef,
leading to a peak of the integral  $\Omega_{xy}^x$.
In contrast,  
  $\sigma_{xy}^z$ is limited, 
leading to a slightly canted spin orientation
in the $xz$-plane. 
The magnitude of SBC was dealt with 
a logarithm function as Eq.~(52) in Ref.~\cite{Qiao2018Dec} for simplicity.
The dynamical stability 
has been validated by its phonon dispersion 
in Supplementary Fig. 7.

\subsection*{High spin Hall conductivity}

Although only a few materials have unconventional SHE components, 
the conventional component \xyz is allowed for all the materials. 
The known \xyz values in 2D materials 
are limited, for instance, 
MoS$_2$ monolayer only presents 
$\sigma_{xy}^z\approx0.2$~\units with doping~\cite{Feng2012Oct}.
To find high values for \xyz in metals, 
SHC calculations are performed on all 210 metals
including semi-metals, 
where 15 materials display exceptional $|\sigma_{xy}^z| > 2$~\units, 
see details in Supplementary Sec. 2. 
Unfortunately, most of them are unstable 
according to phonon dispersions.   
\Cref{tab:metal_shc}  lists 3 materials 
whose mechanical stabilities have been validated by
positive phonon dispersions, 
details are given in 
Supplementary Sec. 3.  

\begin{figure}[tb]
  \includegraphics{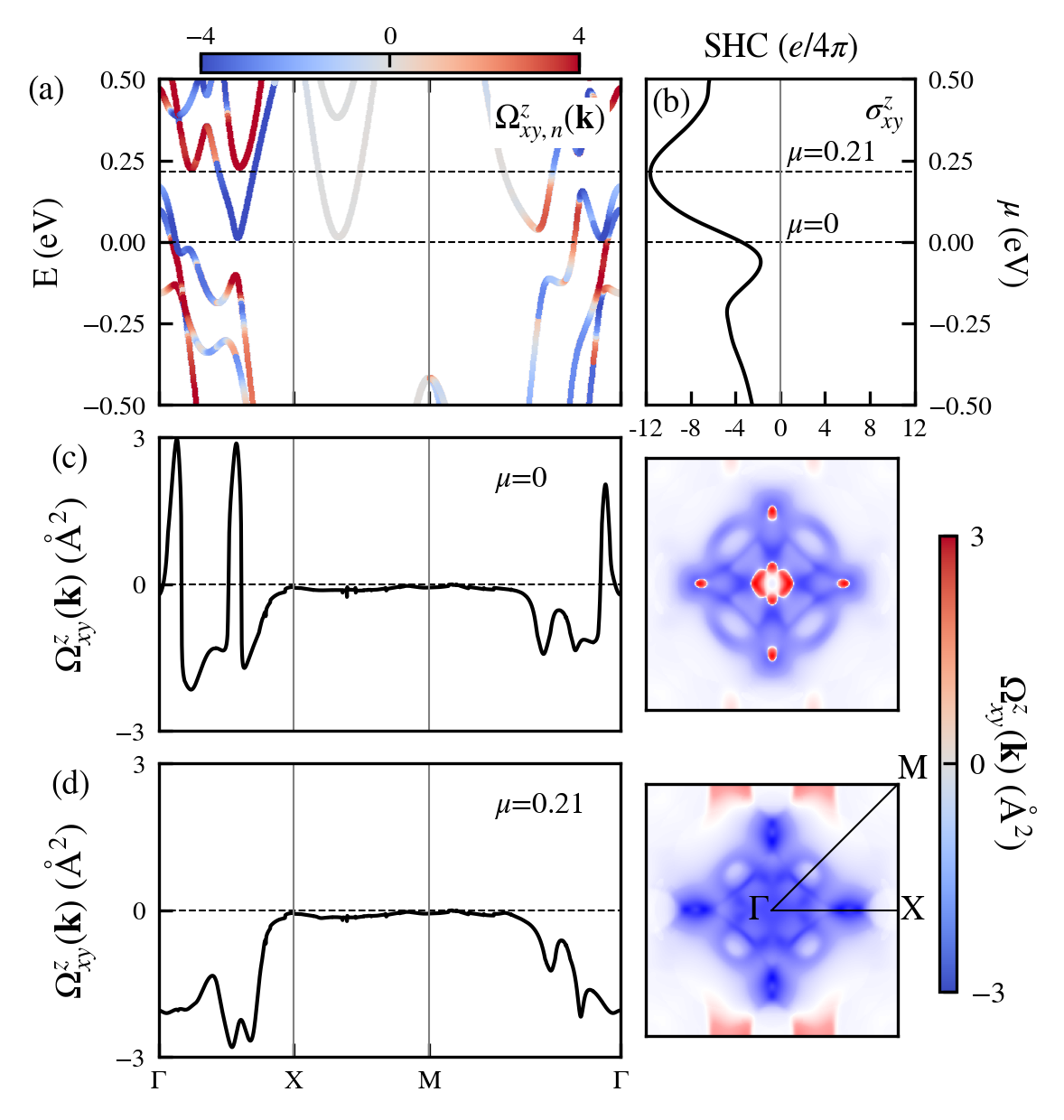}
  \caption{ 
  \textbf{High spin Hall conductivity in \TS.}
  (a) Band structure projected by spin Berry curvature, 
  (b) $\mu$-dependent room-temperature spin Hall conductivity \xyz,
  and   
  $\mathbf{k}$-resolved spin Berry curvatures respectively for  (c) $\mu=0$ and
  (d) $\mu=0.21$~eV
  of \TS monolayer. 
  } 
  \label{fig:Fig2}   
\end{figure}

\TS monolayer presents
the highest value \xyz~=~$-$3.4~\units. 
Considering that \TS is composed of heavy elements, 
its high SHC is attributed to strong SOC
which opens up an anti-crossing band gap around \ef. 
Figures \ref{fig:Fig2}(a) and \ref{fig:Fig2}(b) show
the bands of \TS projected by SBC
and energy-dependent \xyz. 
For $\mu=0$, 
the positive and negative SBCs compete as shown in  \cref{fig:Fig2}(c).
The SHC of \TS can be further enhanced by tuning 
the \ef position.  
For $\mu=0.21$~eV, 
the negative SBC is dominant in the whole BZ as shown in  \cref{fig:Fig2}(d), 
maximizing the \xyz amplitude to be $-$12~\units. 
This finding suggests that 
doping is an effective method to increase SHC in \TS. 

Apart from \TS, 
\SnP and \YB 
are also found to exhibit high \xyz values 
$-$2.7 and 2.4~\units, respectively. 
Note that the high-symmetry \SnP monolayer 
reported by the {\sc{MC2D}} database 
presents   imaginary phonon frequency, 
thus its structure was relaxed without 
preserving any symmetry, leading to the P1 space group.
It is found that
although in a broken-symmetry regime, 
the relaxed structure is very close 
to the original structure, 
namely, the movement of every atom 
is less than $10^{-3}$~\AA.  
Still, this low-symmetry \SnP monolayer 
with positive phonon dispersion is employed to study SHC. 
Besides, 
the possible magnetism in \YB is 
excluded by the negligible magnetic anisotropic energy.
Additional details are given in 
Supplementary Sec. 3.

\subsection*{High spin Hall ratio}

\begin{figure}[tb]
  \includegraphics{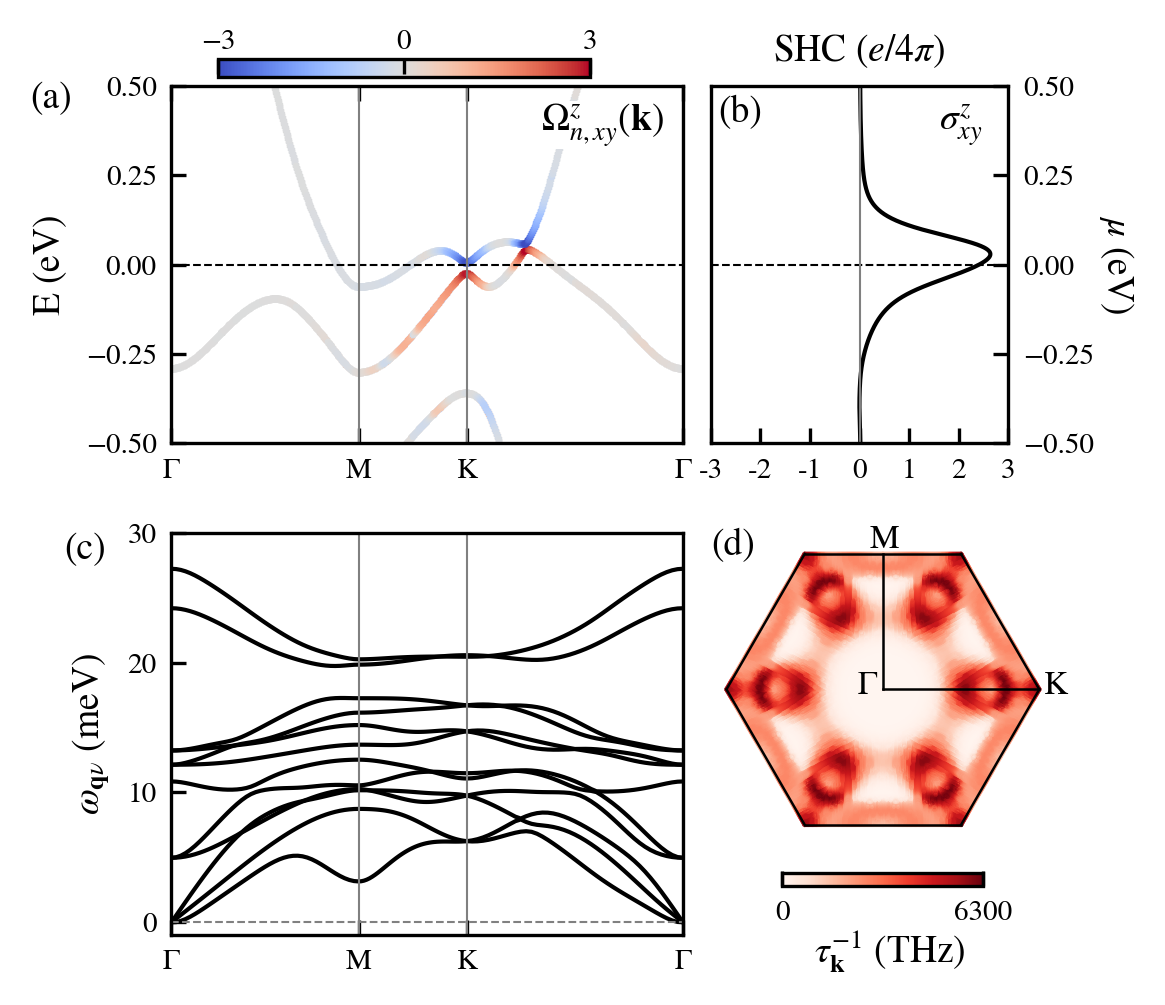}
  \caption{
  \textbf{High spin Hall ratio in \YB.}
  (a)~Bands projected by spin Berry curvature,
  (b)~$\mu$-dependent room-temperature spin Hall conductivity \xyz, 
  (c)~phonon dispersion,
  and 
  (d)~$\bf k$-resolved electron-phonon scattering rate of \YB.
   }
  \label{fig:Fig3}  
\end{figure}

The conversion efficiency from charge current to spin current via the spin Hall
    effect is evaluated by the spin Hall ratio, 
    defined as  $ \xi = \frac{2e}{\hbar} \big|\frac{\sigma_{xy}^z}{\sigma_{xx}} \big|$~\cite{Zhou2024Oct}. 
The semi-metal WTe$_2$ has been reported 
for an efficient conversion \cite{Zhou2019Feb}
since its resistive conductivity \xx is reduced by 
limited carrier around \ef. 
Remarkably, \cref{tab:metal_shc}
includes \YB which is also a semi-metal with high \xyz. 
\Cref{fig:Fig3}(a) shows that 
\YB has a small energy overlap between the conduction band and the valence band.
The SOC-induced anti-crossing gaps 
separate the negative and positive SBCs perfectly, 
leading to a \xyz peak around \ef as given in \cref{fig:Fig3}(b). 
The dynamical stability of \YB is validated by 
the positive phonon dispersion as shown in \cref{fig:Fig3}(c). 
Our finding
suggests that \YB could be a promising candidate for high SHR.

Calculation of charge conductivity 
is required to compute the SHR in  \YB.
For an intrinsic material, 
the electron-phonon coupling (EPC) is a fundamental interaction that affects 
the scattering rate, 
which can be accurately calculated using $ab~initio$ approach~\cite{Giustino2017Feb}. 
The phonon-limited charge conductivity \xx is given by~\cite{Ponce2018Mar} 
  \begin{align}\label{eq:sigma} \sigma_{xx} =
    \frac{-e}{S^{\mathrm{uc}}} \sum_n \int \frac{\mathrm{d}^2
      \mathbf{k}}{\Omega^{\mathrm{BZ}}} v_{n\mathbf{k}x} \partial_{E_{x}}
    f_{n\mathbf{k}},\end{align} where $S^{\rm uc}$ is the unit cell area, $\Omega^{\rm BZ}$ is the BZ area, and $v_{n\mathbf{k}x} = \hbar^{-1} \partial
    \varepsilon_{n\mathbf{k}}/\partial k_{x}$ is the band velocity. 
The linear variation of the electronic occupation function $f_{n\mathbf{k}}$ in
  response to ${E_x}$, $\partial_{E_{x}}f_{n\mathbf{k}} $, can be
  obtained by solving the Boltzmann transport equation~\cite{Ponce2020Feb}
   which induces the scattering rate given by
  \begin{multline}\label{eq:scattering_rate} \tau_{n\mathbf{k}}^{-1} =
  \frac{2\pi}{\hbar} \sum_{m\nu} \!
\int\! \frac{\mathrm{d}^2 \mathbf{q}}{\Omega^{\text{BZ}}} | g_{mn\nu}(\mathbf{k,q})|^2
   \big[ (n_{\mathbf{q}\nu} +1 - f_{m\mathbf{k+q}}^0) \delta( \varepsilon_{n\mathbf{k}} - \varepsilon_{m\mathbf{k+q}}   -  \hbar \omega_{\mathbf{q}\nu})\\
  + (n_{\mathbf{q}\nu}  + f_{m\mathbf{k+q}}^0 )\delta(\varepsilon_{n\mathbf{k}} - \varepsilon_{m\mathbf{k+q}} + \hbar \omega_{\mathbf{q}\nu}) \big],
\end{multline}
where $g_{mn\nu}(\mathbf{k,q})$ is the electron-phonon matrix element with  phonon frequency $\omega_{\mathbf{q}\nu}$. 
The $\bf k$-resolved effective scattering rate, 
defined as  
$ \tau^{-1}_{\bf{k}} = \sum_n
    \frac{-\partial f_{n \bf{k}}^0}{\partial \varepsilon_{n \bf{k}}} {\tau^{-1}_{n \bf{k}}}$,
is given in \cref{fig:Fig3}(d). 
Strong scattering occurs around the K points 
where   \ef crosses  the  anti-crossing gap.
The small bandgap facilitates scattering between electron states, 
enhancing EPC and scattering rates. 
Besides, the pattern of \cref{fig:Fig3}(d) 
is determined by structural symmetry
which also forces $\sigma_{xx} = \sigma_{yy}  $ in \YB. 

The SHR of \YB can be obtained from the calculated 
SHC and charge conductivity. 
\YB  presents an SHC of $\sigma_{xy}^{z} = 2.4~e/4\pi$,
and a charge conductivity of $\sigma_{xx} = 6.1~e^2/h$. 
Defined as $ \xi = \frac{2e}{\hbar} \big|\frac{\sigma_{xy}^z}{\sigma_{xx}} \big|$, 
the room-temperature SHR in \YB is calculated to be $ \xi =0.4$, 
much higher than SHRs $\sim0.01$ in typical heavy metals~\cite{Tao2018Jun}.
Our results demonstrate that 
attributed to anti-crossings in semi-metal bands,
\YB is a promising candidate  for efficient charge-to-spin conversion.

\subsection*{Quantized spin Hall conductivity}

As the last part of this work, gapped materials
are investigated.
It is instructive to compute the SHC of materials to discover topological insulators
which possess a perfect SHC plateau inside their bandgap. 
A typical QSHI  presents 
a quantized SHC with amplitude of 
$\sigma_{xy}^z = n~e/2\pi$
where  $n$  is an integer
\cite{Kane2005Nov}. 
However,
in practice, the SHC plateau is not always well quantized;
namely, some $\mathbb{Z}_2=1$ QSHIs manifest significant deviations from their
expected quantized value~\cite{Costa2021Apr}.

To clarify the mechanism of quantized SHC, 
we calculated the \xyz of 216 insulators 
with $T = 0$ K and $\eta = 5$~meV. 
We find 8 materials with SHC plateau in the bandgap. 
Among these materials,  
Bi$_2$, Hf$_2$Br$_2$, CuLi$_2$As, and Ti$_2$N$_2$I$_2$
have been validated as QSHIs 
with  $\mathbb{Z}_2=1$~\cite{Grassano2023Sep}. 
The other 4 materials are 
Li$_2$Tl$_2$, Hg$_3$S$_2$,  Hg$_4$O$_2$, and  In$_2$Bi$_2$
which possess soft phonon modes.
Thus we only list the results of the former 4 materials in 
\cref{tab:TIs}.
All 4 materials present a non-zero SHC plateau 
inside the bandgap. 
However, the SHC amplitudes vary
and  the $\beta$-Bi$_2$ monolayer shows  \xyz=~0.5~\unitqs 
while \CLA presents a quantized SHC of \xyz=~$1.0$~\unitqs. 

\begin{figure}[tb]
  \includegraphics{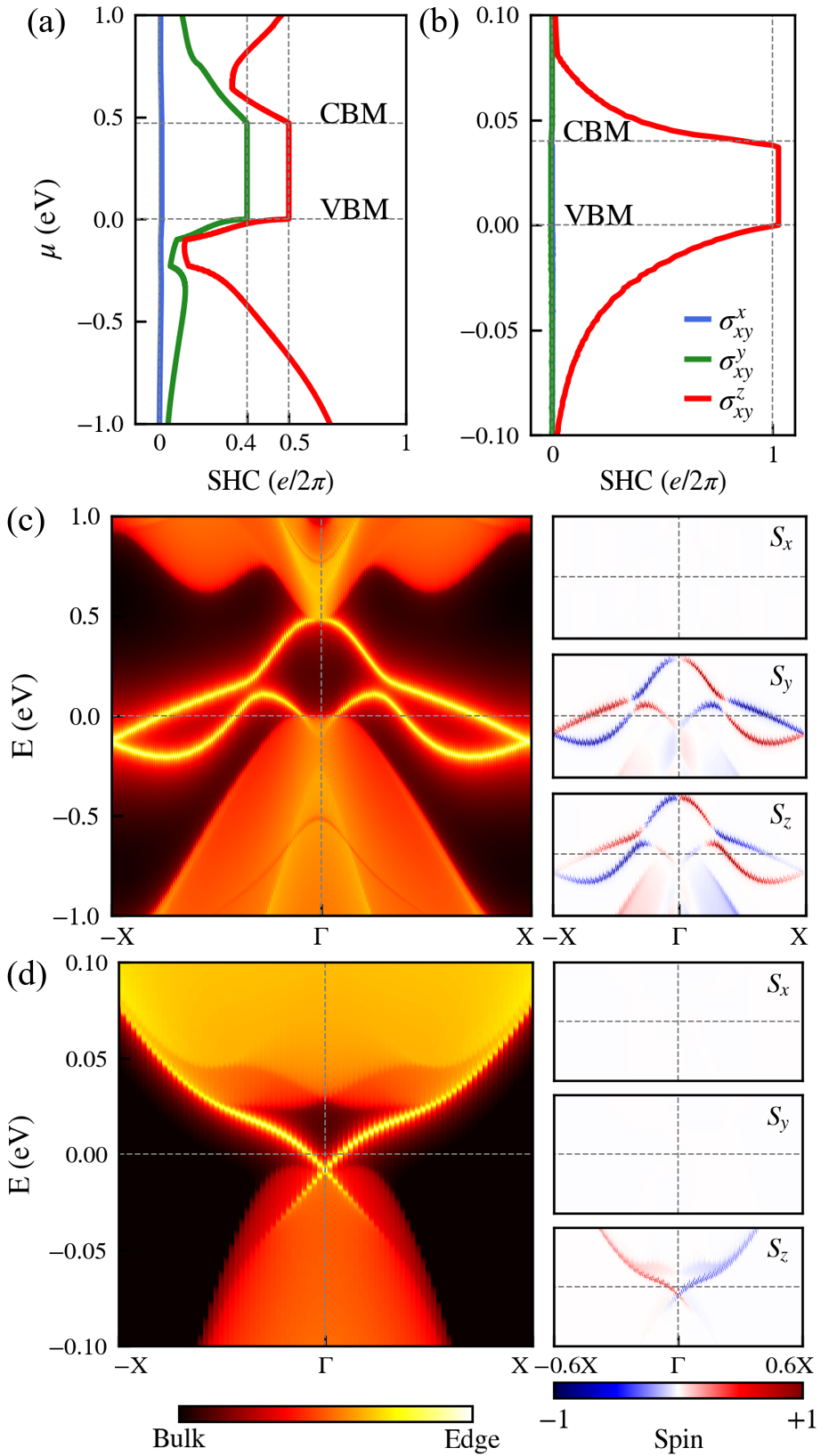}
  \caption{
  \textbf{Spin Hall conductivities and edge states in quantum spin Hall insulators.}
  $\mu$-dependent spin Hall conductivity components of (a)~Bi$_2$ and (b)~\CLA at $T=0$~K.
  Edge states and spin projections of (c)~Bi$_2$ and (d)~\CLA.
  The conduction band minimum is abbreviated as CBM, 
  valence band maximum is abbreviated as VBM.
   }
  \label{fig:Fig4}  
\end{figure}

The potentially quantized SHC can be interpreted by spin conservation. 
Symmetry analysis~\cite{Roy2022Apr} illustrates that 
in the $\beta$-Bi$_2$ monolayer, 
SHC components of \xyy and \xyz are allowed
while \xyx is prohibited.
Indeed, \cref{fig:Fig4}(a) shows
non-zero \xyy and \xyz SHC plateaus inside bandgap. 
Besides, 
the edge-state spin of Bi$_2$ is neither conserved nor robust
as shown in \cref{fig:Fig4}(c). 
In contrast, 
\cref{fig:Fig4}(b) shows that only \xyz is allowed in \CLA, 
which presents 
a pair of helical edge states 
with well-conserved  $S_z$  as shown in  \cref{fig:Fig4}(d). 
As a result, we find a quantized SHC 
\xyz=~1.0~\unitqs in \CLA.
Besides, with the same space group as Bi$_2$, 
Hf$_2$Br$_2$ also presents
a deviation from quantized SHC 
due to non-zero \xyy and non-zero $S_y$ on edge states.
Ti$_2$N$_2$I$_2$ presents similar behaviors as \CLA.
It can be concluded that 
the non-conserved $S_z$ is destructive for quantized SHC. 
Calculations of SHC and spin-projected edge states are necessary 
to predict the value of SHC  plateau in QSHI.

In conclusion, leveraging the {\sc{MC2D}} database, 
we implemented an automated Wannierization 
based on orbital information 
from the pseudopotentials
for 426 non-magnetic monolayers
including  210 metals and 216 insulators. 
The quality of Wannierizations has been 
validated by comparing DFT bands with Wannier bands, 
and the Wannier Hamiltonians are provided to the community. 
Exotic spin Hall effects are discovered  in diversified materials.
For metals, it is found that \YCI displays canted spin Hall effect 
due to its low-symmetry structure. 
Besides,  metallic candidates with high SHC values are listed, 
indicating \TS as the optimal candidate. 
Efficient charge-to-spin conversion is reported in 
the semi-metal \YB.  
For insulators, 
SHC plateaus are revealed in 
quantum spin Hall insulators, 
highlighting the importance of 
 conserved $S_z$ for quantized SHC. 
This comprehensive work
enhances the understanding of spin Hall effect in 2D systems, 
suggesting promising materials for advanced spintronic devices. 
Moreover, 
the provided Wannier database could be further employed to explore 
additional spin-orbitronic phenomena.

\section*{Methods}

\subsection*{DFT calculations}
DFT  calculations 
were performed using the
{\sc Quantum ESPRESSO} package~\cite{Giannozzi_2017} 
within the Perdew-Burke-Ernzerhof (PBE)
parametrization of generalized gradient approximation
(GGA).
{\sc{AiiDA}}~\cite{Huber2020Sep} was employed to calculate  
  426 non-magnetic monolayers from {\sc{MC2D}} database~\cite{MC2D1,MC2D2}
  with a $\bf k$-mesh density of 0.15~\AA$^{-1}$,
  Marzari-Vanderbilt smearing~\cite{Marzari1999Apr}  of 0.01~Ry, 
and 2D Coulomb truncation scheme~\cite{Sohier2017Aug}. 
Although 
the {\sc{SSSP}} library~\cite{Prandini2018} 
was used in {\sc{MC2D}} database, 
it only includes scalar-relativistic pseudopotentials.
To investigate SOC effect, 
our work employed 
{\sc{PseudoDojo}} library~\cite{vanSetten2018May} which
includes both scalar-relativistic and fully relativistic pseudopotentials. 
The fully relativistic pseudopotentials 
were applied to {\sc{MC2D}} materials without relaxation.
After screening out 8  candidates 
with superior SHC properties, 
the scalar-relativistic {\sc{PseudoDojo}} pseudopotentials were used
to relax structures and calculate phonon dispersions.
For the relaxed 8 structures, 
the fully relativistic {\sc{PseudoDojo}} pseudopotentials were applied 
to the DFT calculations and Wannierizations.  

\subsection*{Wannierizations}
In our work, 
the initial guess for projector is given by the
hydrogenic orbital with hints from
  {\sc{PseudoDojo}}~\cite{semicore}.
The semicore orbitals corresponding to deep bands are excluded from
  projections.
The maximum of the frozen window is set as \ef+~2~eV
(\ef is the conduction band minimum for insulators)
which is numerically verified to be the optimal value
as shown in  Supplementary Fig. 1(a).
To provide a high-quality Wannier Hamiltonian database, 
failed cases (42/426)
have been manually corrected by
    tuning the frozen window, 
    increasing the number of electronic states,
    or modifying initial projections.

\subsection*{Transport calculations}
The spin Hall conductivity was calculated with the \textsc{Wannier90}
  package~\cite{Pizzi_2020,Qiao2018Dec}
  with a  fine $\mathbf{k}$-mesh density of 0.005~\AA$^{-1}$ for the Wannier interpolation.
A Fermi-Dirac distribution function of $T=300$~K and a broadening of 2~meV were considered
if not specified. 
The charge conductivity  was calculated using the
  \textsc{EPW} package~\cite{Ponce2016, Lee2023Feb}.
A coarse $12\times 12\times1$ $\mathbf{k}$/$\mathbf{q}$-grid was
interpolated on a fine $\mathbf{k}$/$\mathbf{q}$-grid of  $360\times 360 \times1$. 
The electron wavefunctions  and dynamical matrices 
were  calculated 
using fully and scalar relativistic pseudopotentials, 
respectively~\cite{Zhou2024Sep}. 
An adaptive smearing~\cite{Ponce2021Oct}, 
  a phonon frequency cutoff of $1~\mathrm{cm}^{-1}$ 
  and a 300~K temperature are used. 
Edge states of topological insulators
were calculated by WannierTools~\cite{Wu2018Mar}.

\section*{Data availability}
Details including electronic structures and phonon dispersions 
of promising materials investigated in this work,
as well as fully relativistic bands 
of 426 monolayers,
are given in the Supplementary Material. 
The input and output files, Wannier Hamiltonians, 
and bands data of promising materials and 426 monolayers
are provided on Materials Cloud Archive~\cite{MCA}.


\section*{Acknowledgements}
S.~P.
acknowledges the support from the Fonds de la Recherche Scientifique de Belgique (\frs-FNRS).
J.~Z.
and {J.-C.~C.}
acknowledge financial support from the F\'ed\'eration Wallonie-Bruxelles through the ARC Grant ``DREAMS'' (No. 21/26-116), from the EOS project ``CONNECT'' (No. 40007563), and from the Belgium \frs-FNRS through the research project (No.~T.029.22F).
Computational resources have been provided by the PRACE/EuroHPC award granting access
  to MareNostrum5 at Barcelona Supercomputing Center (BSC), Spain and Discoverer
  at SofiaTech, Bulgaria (OptoSpin project ID.
2020225411), and by the Consortium des Équipements de Calcul Intensif (C\'ECI), funded by the \frs-FNRS under Grant No. 2.5020.11 and by the Walloon Region.

\section*{Author contributions}
J.~Z. designed the study and performed calculations. S.~P. and {J.-C.~C.} supervised the project. All authors analyzed the results and contributed to writing the paper.

\section*{Competing interests}
The authors declare no competing interests.

\section*{Additional information}
\subsection*{Supplementary information}
The supplementary material has been attached. 

\clearpage

\clearpage

\clearpage

\clearpage

\clearpage

\renewcommand{\arraystretch}{1.2}
\newcolumntype{d}[1]{D{.}{.}{#1}}
\begin{table}[b!]
  \centering
  \caption{\textbf{Stable metals with high conventional
  spin Hall conductivity ${\sigma_{xy}^z}$.}}
    \begin{tabular}{  C{2mm} C{2.5cm} C{2.5cm} C{2.5cm}   C{2mm}  }
      \hline \hline 
      & Material     & Space group    & \xyz~(\units) & \\  \hline   
      & Ta$_4$Se$_2$        & P4/nmm    & $-$3.4      & \\
      & Sn$_2$P             & P1        & $-$2.7    & \\ 
      & Y$_2$Br$_2$         & P-3m1     & 2.4       & \\ 
      \hline  \hline
  \end{tabular}
  \label{tab:metal_shc}
\end{table}

\begin{table}[bth]
  \centering
  \caption{\textbf{List of quantum spin Hall insulators.}
  SG is the abbreviation of space group,
  \xyz denotes the plateau value in band gap,
  and 
  $S_z$ illustrates whether the out-of-plane  spin is conserved or not. 
}
    \begin{tabular}{  C{2cm} C{2.5cm}   C{1.6cm} C{2.2cm} C{1.1cm}}
      \hline \hline 
      Monolayer  & $E_{\rm gap}$ (meV) & SG & \xyz (${e}/{2\pi}$)     & ${S_z}$  \\ \hline 
      Bi$_2$            & 471    & P-3m1    & $ ~0.5$   & no       \\ 
      Hf$_2$Br$_2$      & 69     & P-3m1    & $ -0.9$   & no       \\ 
      CuLi$_2$As        & 41     & P-6m2    & $ ~1.0$   & yes      \\ 
      Ti$_2$N$_2$I$_2$  & 19     & Pmmn     & $ -1.0$   & yes      \\  
      \hline  \hline
  \end{tabular}
  \label{tab:TIs}
\end{table}

\end{document}